\documentclass[12pt]{article}
\input epsfig.sty

\font\grande=cmr9.5 scaled \magstep4
\font\medio=cmr9.5 scaled \magstep2
\outer\def\beginsection#1\par{\medbreak\bigskip
      \message{#1}\leftline{\bf#1}\nobreak\medskip
\vskip-\parskip
      \noindent}
\begin{document}
\bibliographystyle {unsrt}

\titlepage

\begin{flushright}
CERN-PH-TH/2014-096
\end{flushright}

\vspace{10mm}
\begin{center}
{\grande Cosmic backgrounds of relic gravitons}\\ 
\vspace{5mm}
{\grande and their absolute normalization}\\
\vspace{10mm}
 Massimo Giovannini 
 \footnote{Electronic address: massimo.giovannini@cern.ch} \\
\vspace{0.5cm}
{{\sl Department of Physics, Theory Division, CERN, 1211 Geneva 23, Switzerland }}\\
\vspace{1cm}
{{\sl INFN, Section of Milan-Bicocca, 20126 Milan, Italy}}
\vspace*{1.5cm}

\end{center}

\centerline{\medio  Abstract}
\vspace{5mm}
Provided the consistency relations are not violated, the recent Bicep2 observations pin down 
the absolute normalization, the spectral slope and the maximal frequency of the cosmic graviton background produced during inflation.
The properly normalized spectra are hereby computed from the lowest frequencies (of the order of the present Hubble rate) 
up to the highest frequency range in the GHz region.  Deviations from the conventional paradigm cannot be excluded and are examined by allowing for different physical possibilities including, in particular, a running of the tensor spectral index, an explicit breaking 
of the consistency relations and a spike in the high-frequency tail of the spectrum coming either from a post-inflationary phase dominated by a stiff fluid of from the contribution of waterfall fields in a hybrid inflationary context. 
The direct determinations of the tensor to scalar ratio at low frequencies, 
if confirmed by the forthcoming observations, will also affect and constrain the high-frequencies uncertainties. 
The limits on the cosmic graviton backgrounds coming from wide-band interferometers (such as LIGO/Virgo, LISA and BBO/DECIGO) 
together with a more accurate scrutiny of the tensor B mode polarization at low frequencies will set direct bounds on the post-inflationary evolution and on other unconventional completions of the standard lore. 
\noindent

\vspace{5mm}
\vfill

\newpage
\renewcommand{\theequation}{1.\arabic{equation}}
\setcounter{equation}{0}
\section{Bicep2 observations and relic gravitons}
\label{sec1}
The Bicep2 experiment \cite{bicep2} observed recently the polarization of the Cosmic Microwave Background 
(CMB in what follows) and reported the detection of a B mode component that can be well fit within the standard $\Lambda$CDM scenario\footnote{In the $\Lambda$CDM scenario the $\Lambda$ qualifies the dark energy component while the CDM stands for the dark matter component.} supplemented by 
a tensor-to-scalar-ratio $r_{T}=0.2^{+0.07}_{-0.05}$.  The first detection of a B mode polarization, coming from the lensing of the CMB anisotropies, has been published some time ago by the South Pole Telescope \cite{SPTpol}.
The lensed $\Lambda$CDM paradigm can be further complemented by $r_{T}$, i.e. the ratio of the tensor and scalar power spectra at a conventional pivot scale\footnote{ The WMAP collaboration \cite{WMAP1,WMAP9}
consistently chooses $k_{p}= 0.002\, \, \mathrm{Mpc}^{-1}$. The Bicep2 collaboration uses $k_{p} =0.05\, \mathrm{Mpc}^{-1}$ \cite{bicep2}.
The  Planck collaboration \cite{planck} assigns the scalar power spectra of curvature perturbations ${\mathcal P}_{{\mathcal R}}$ 
at $k_{p} =0.05\, \mathrm{Mpc}^{-1}$ while the tensor to scalar ratio $r_{T}$ is assigned at $k_{p} =0.002\, \mathrm{Mpc}^{-1}$.}
$k_{p}$:
\begin{equation}
r_{T} = \frac{{\mathcal A}_{T}}{{\mathcal A}_{R}};
\label{r1}
\end{equation}
where ${\mathcal A}_{T}$ and ${\mathcal A}_{{\mathcal R}}$ denote, respectively, the amplitudes of the tensor and scalar power spectra
\begin{equation}
{\mathcal P}_{T}(k) = {\mathcal A}_{T} \biggl(\frac{k}{k_{p}}\biggr)^{n_{T}}, \qquad {\mathcal P}_{{\mathcal R}}(k) = {\mathcal A}_{{\mathcal R}} \biggl(\frac{k}{k_{p}}\biggr)^{n_{s}-1}
\label{r2}
\end{equation}
and  $n_{s}$ and $n_{T}$ are the corresponding spectral indices. For an explicit determination of $r_{T}$, the Bicep2 collaboration confronts the observations of the B mode power spectra with the 
tensor extension of the lensed $\Lambda$CDM paradigm. Such a model is minimal insofar as it involves a single supplementary parameter,
i.e. $r_{T}$ that is related to the tensor spectral index and 
to the slow roll parameter $\epsilon$ by the so called consistency relation:
\begin{equation}
r_{T}= 16\,\epsilon = - 8 n_{T},
\label{r3}
\end{equation}
where $\epsilon = - \dot{H}/H^2$ denotes the slow roll parameter measuring the decrease of the Hubble rate during 
inflation\footnote{In  the present notations $H= \dot{a}/a$ is the Hubble rate and the overdot denotes a derivation with 
respect to the cosmic time coordinate. }. The consistency relation (\ref{r3}) can be violated by the initial conditions of the 
tensor modes. In the present investigation we shall first examine the case where the consistency relations 
are enforced but we shall then deviate from this possibility and analyze more general situations that are less conventional but still 
not ruled out by the Bicep2 observations.

In the light of the lensed $\Lambda$CDM scenario supplemented by tensors, the results of Ref. \cite{bicep2} imply that
 the value $r_{T}=0$ is disfavoured at more than  $5\sigma$.  It is plausible that the actual 
primordial component of $r_{T}$ will be slightly smaller than $0.2$ 
even if  various tests were performed on the data to eliminate systematic effects 
and other contaminations from galactic synchrotron and from polarized 
dust emissions. After subtraction of some purported foregrounds the values of $r_{T}$ may get closer to the Planck limits \cite{planck,emission} and 
imply, presumably, $r_{T} = 0.16^{+0.06}_{-0.05}$. 
For the aims of this paper we shall preferentially consider $r_{T} =0.2$ as fiducial value but also allow for slightly smaller
values. The range of values of $r_{T}$ will be, in practice, from $0.16$ to $0.2$; possible ambiguities 
stemming from the different conventional choices of the pivot scale shall be briefly addressed in section \ref{sec3}.

There are some who suggest that the measured value of $r_{T}$  are in moderate tension with other satellite data \cite{WMAP9,planck}. The Bicep2 
observations  do not contradict the WMAP 9-yr upper limits on $r_{T}$. Conversely the Planck bound \cite{planck}, obtained from the temperature anisotropies under the assumption of a constant scalar spectral index $n_s$,
stipulates that $r_{T} < 0.11$.  This constraint is not obtained by the Planck collaboration without
assuming some of the E-mode polarization observables that are taken from the WMAP 9-yr measurements.
The Planck limit can be relaxed to $r_{T} < 0.26$ if the scalar spectral index is scale dependent. 
If $n_s$ varies with $k$ the scalar power spectrum can be parametrized as in Eq. (\ref{r2}) 
but with the spectral indices given by\footnote{The derivation of the consistency 
relations, of the well known spectral relations assumed in Eqs. (\ref{r1})--(\ref{r3}) and of the basic formulas of slow-roll dynamics can be found in different 
books \cite{wein,mg} and, with slightly different notations, in earlier references (see e.g. \cite{Ly}).}
\begin{eqnarray}
n_{s} &=& 1  - 6 \epsilon + 2 \overline{\eta} +\frac{1}{2}\alpha_{s} \ln{(k/k_{p})},
\label{break2}\\
n_{T} &=& - 2 \epsilon + \frac{1}{2}\alpha_{T} \ln{(k/k_{p})},
\label{break3}
\end{eqnarray}
where $\overline{\eta}=   \overline{M}^2_{\mathrm{P}} V_{\varphi\varphi}/V$ and $\xi^2 = \overline{M}_{P}^2 V_{,\varphi}\,\, V_{,\, \varphi\, \varphi\,\varphi}/V^2$ denote the two supplementary slow-roll parameters expressed as derivatives of the inflaton potential.
In Eq. (\ref{break3})  $\alpha_{s}$ and $\alpha_{T}$ denote, respectively, the scalar and the tensor running:
\begin{equation}
\alpha_{s} = \frac{1}{2} r_{T} (n_{s} -1) + \frac{3}{32} r_{T}^2 - 2 \xi^2, \quad 
\alpha_{T} = \frac{r_{T}}{8} \bigg[ (n_{s}-1) + \frac{r_{T}}{8} \biggr].
\label{break4}
\end{equation} 
To relax the Planck we must require
$\alpha_s \simeq - 0.02$ so that the temperature data can be  reconciled with the $r_{T} =0.2$ value of the Bicep2 data. 
Unfortunately, Eq. (\ref{break4}) would imply that $\alpha_{s}$ is much lower 
in inflationary modes and more ${\mathcal O}(10^{-4})$ (for  $n_s = 0.96$ and  $r=0.2$) and always with negative sign.

Barring for some possible reduction due to foreground subtraction, in the present paper we shall not dwell about the possibility of reconciling the various upper limits but rather acknowledge that the value $r_{T}=0$ is ruled out and that the tensor to scalar ratio is ${\mathcal O}(0.2)$.  With these specifications, the Bicep2 observations, if taken at face value,  imply an absolute normalization of the relic graviton background 
for typical frequencies comparable with the pivot frequency $\nu_{p}$, i.e. 
\begin{equation}
\nu_{\mathrm{p}} = 3.092\,\biggl(\frac{k_{p}}{0.002\,\,\mathrm{Mpc}^{-1}}\biggr) \,  10^{-18} \,\, \mathrm{Hz}.
\label{fr1}
\end{equation}
Recalling the comoving angular diameter distance to last scattering $d_{A}(z_{*}) = (14029 \pm 119) \, \mathrm{Mpc}$ and to 
equality $d_{A}(z_{eq}) = (14194 \pm 117) \, \mathrm{Mpc}$ \cite{WMAP9} we have, within an order of magnitude,
that $\nu_{\mathrm{p}} \simeq 1/d_{A}(z_{*})\simeq 1/d_{A}(z_{eq}) \simeq H_{0}$ where 
$H_{0} = 2.26 \times 10^{-18} \, (h_{0}/0.7) \,\, \mathrm{Hz}$ denotes the Hubble rate.

The knowledge of the normalization of the cosmic graviton background and of its slope can be hardly underestimatd. In the 
concordance paradigm the results of Ref. \cite{bicep2} not only provide a robust low-frequency 
normalization but also determine the high frequency tail of the spectrum. 
In spite of its success, the minimal $\Lambda$CDM with tensors may not be the end of the story. It is then wise to bear in mind the possibility that there is some running of the tensor spectral index (i.e. $\alpha_{T}\neq 0$ in Eq. 
(\ref{break4})) or that the consistency relations may not be satisfied because $r_{T}$ and $n_{T}$ are independently assigned. In both situations the high-frequency 
regime (between the MHz and the GHz) is affected and it may become even more uncertain
because the precise nature of the post-inflationary evolution 
is not fully determined in the concordance paradigm \cite{mg1}. There could be a phase of prolonged reheating;
the sound speed of the plasma 
may be stiffer than radiation as suggested long ago by Zeldovich \cite{zel1} (see also \cite{mg1,pee}). A further possibility is that the graviton background has a secondary component  such as the one induced by the anisotropic stress created by some 
spectator or waterfall field (see e.g. \cite{wat1,wat2}). In both cases the spectral energy density is enhanced at high frequencies as argued in the past \cite{mg2}.

It seems therefore timely to discuss with a certain degree of accuracy the cosmic backgrounds of relic gravitons
both at low and at higher frequencies in the light of the current developments stemming from the analysis 
of the CMB polarization. This analysis can be useful for the higher frequency experiments aimed at a direct detection of relic gravitons such as the 
various wide-band interferometers, i.e. the LISA project \cite{lisa}, the BBO/DECIGO \cite{BBODECIGO} projects or the LIGO/Virgo experiments 
\cite{LIGOS1,LIGOS2,LIGOS3,virgoligo,LIGO4}. The layout of this paper is the following.
In section \ref{sec2} we will introduce the spectra of cosmic gravitons and the typical frequency scales of the problem.
In section \ref{sec3} the cosmic backgrounds of relic gravitons will be computed in the case of the conventional inflationary models. In section \ref{sec4} we shall analyze the potential high-frequency uncertainties. 
The concluding remarks shall be collected in section \ref{sec5}.

\renewcommand{\theequation}{2.\arabic{equation}}
\setcounter{equation}{0}
\section{The spectra and their typical frequency scales}
\label{sec2}
\subsection{Preliminaries} 
Let us start by defining three mutually orthogonal directions: $\hat{k}_{i} = k_{i}/|\vec{k}|$,  $\hat{m}_{i} = m_{i}/|\vec{m}|$ and $\hat{n} =n_{i}/|\vec{n}|$. The two polarizations of the gravitons in a conformally flat background geometry are:
 \begin{equation}
 e_{ij}^{(\oplus)}(\hat{k}) = (\hat{m}_{i} \hat{m}_{j} - \hat{n}_{i} \hat{n}_{j}), \qquad 
 e_{ij}^{(\otimes)}(\hat{k}) = (\hat{m}_{i} \hat{n}_{j} + \hat{n}_{i} \hat{m}_{j}),
 \label{ST0}
 \end{equation}
 where $\hat{k}$ is oriented along the direction of propagation of the wave. It follows from Eq. (\ref{ST0}) that $e_{ij}^{(\lambda)}\,e_{ij}^{(\lambda')} = 2 \delta_{\lambda\lambda'}$ while the sum over the polarizations gives:
\begin{equation}
\sum_{\lambda} e_{ij}^{(\lambda)}(\hat{k}) \, e_{m n}^{(\lambda)}(\hat{k}) = \biggl[p_{m i}(\hat{k}) p_{n j}(\hat{k}) + p_{m j}(\hat{k}) p_{n i}(\hat{k}) - p_{i j}(\hat{k}) p_{m n}(\hat{k}) \biggr];
\label{ST0B} 
\end{equation}
where $p_{ij}(\hat{k}) = (\delta_{i j} - \hat{k}_{i} \hat{k}_{j})$. Defining the Fourier transform of $h_{ij}(\vec{x},\tau)$ as 
\begin{equation}
h_{ij}(\vec{x},\tau) = \frac{1}{(2\pi)^{3/2}}\, \sum_{\lambda}  \, \int d^{3} k \, h_{ij}(\vec{k},\tau)\,\, e^{- i \vec{k}\cdot\vec{x}}, \qquad 
h_{ij}(\vec{k},\tau) = \sum_{\lambda= \otimes, \,\oplus} \, e^{(\lambda)}_{ij}(\hat{k}) \, h_{(\lambda)}(\vec{k},\tau),
\label{St0C}
\end{equation}
the tensor power spectrum ${\mathcal P}_{T}(k,\tau)$ determines the two-point function at equal times:
\begin{equation}
\langle h_{ij}(\vec{k},\tau) \, h_{mn}(\vec{p},\tau) \rangle = \frac{2\pi^2}{k^3} {\mathcal P}_{T}(k,\tau) \, {\mathcal S}_{ijmn}(\hat{k}) \delta^{(3)}(\vec{k} +\vec{p}),
\label{ST1}
\end{equation}
where, recalling Eqs. (\ref{ST0}) and (\ref{ST0B}), 
\begin{equation}
{\mathcal S}_{ijmn}(\hat{k}) = \frac{1}{4} \sum_{\lambda} e_{ij}^{(\lambda)}(\hat{k}) \, e_{m n}^{(\lambda)}(\hat{k}).
\label{ST1B}
\end{equation}
Equation (\ref{ST1}) follows the same conventions used when assigning the curvature perturbations
\begin{equation}
\langle {\mathcal R}(\vec{k},\tau) \, {\mathcal R}(\vec{p},\tau) \rangle = \frac{2\pi^2}{k^3} {\mathcal P}_{{\mathcal R}}(k,\tau)  \delta^{(3)}(\vec{k} +\vec{p});
\label{ST1C}
\end{equation}
${\mathcal R}(\vec{x},\tau)$ denotes the curvature perturbations on comoving orthogonal hypersurfaces and it is the quantity customarily 
employed to set the initial conditions of the Einstein-Boltzmann hierarchy in the observational analyses (see e.g. \cite{WMAP1,WMAP9}).
\subsection{Power spectra and spectral energy density}
The equations obeyed by $h_{ij}(\vec{x},\tau)$ follows from the second order action: 
\begin{equation}
S = \frac{1}{8 \ell_{\mathrm{P}}^2} \int d^{4} x \,\,\sqrt{- \overline{g}}\,\,\overline{g}^{\alpha\beta}
\partial_{\alpha} \, h_{i}^{j} \, \partial_{\beta} h_{j}^{i},
\label{action1}
\end{equation}
where $\overline{g}_{\mu\nu}$ denotes a conformally flat background metric\footnote{Greek letters are used to denote four-dimensional indices; lowercase Latin characters denote spatial indices.} that can be written as  $\overline{g}_{\mu\nu} = a^2(\tau)\eta_{\mu\nu}$. Note that in Eq. (\ref{action1}) $\ell_{\mathrm{P}} = \sqrt{8 \pi G} = 1/\overline{M}_{\mathrm{P}}$. In what follows we shall distinguish (especially in the last section) between $\overline{M}_{\mathrm{P}}$ and $M_{\mathrm{P}} = 1.22\times 10^{19} \mathrm{GeV} = \sqrt{8 \pi} \,\,\overline{M}_{\mathrm{P}}$. In the action (\ref{action1}) $h_{ij}$ denote the traceless and divergenceless modes of the geometry written in the form $g_{\mu\nu}(\vec{x},\tau) = 
\overline{g}_{\mu\nu}(\tau) + \delta_{\mathrm{t}}g_{\mu\nu}(\vec{x},\tau)$ where 
$\delta_{\mathrm{t}} g_{i j } = - a^2 \, h_{ij}$ and $\delta g^{i j} = h^{ij}/a^2$.
By taking the functional variation of the action (\ref{action1}) with respect to $h_{i}^{j}$ the equation of motion reads:
\begin{equation}
h_{ij}^{\prime\prime} + 2 {\mathcal H} h_{ij}^{\prime} - \nabla^2 h_{ij} =  - 2 \ell_{P}^2 a^2 \Pi_{ij}(\vec{x},\tau),
\label{mot1A}
\end{equation}
where the prime denotes a derivation with respect to the conformal time 
coordinate $\tau$ and ${\mathcal H} = (\ln{a})' = a\,H$ where $H$ is the Hubble rate. In Eq. (\ref{mot1A})  
the contribution of the transverse and traceless anisotropic stress has been added for convenience. At low frequencies and in the concordance 
paradigm the contribution to $\Pi_{ij}$ is due to the presence of (effectively massless) neutrinos.  At high frequencies the anisotropic 
stress induced by waterfall fields may lead to an enhancement of the spectral energy density \cite{mg2}.

During inflation, when the anisotropic stress is immaterial, $h_{ij}(\vec{x},\tau)$ can be quantized and the corresponding 
field operator can be written as:
\begin{equation}
\hat{h}_{ij}(\vec{x},\tau) = \frac{\sqrt{2} \ell_{\mathrm{P}}}{(2\pi)^{3/2}}\sum_{\lambda} \int \, d^{3} k \,\,e^{(\lambda)}_{ij}(\vec{k})\, [ F_{k,\lambda}(\tau) \hat{a}_{\vec{k}\,\lambda } e^{- i \vec{k} \cdot \vec{x}} + F^{*}_{k,\lambda}(\tau) \hat{a}_{\vec{k}\,\lambda }^{\dagger} e^{ i \vec{k} \cdot \vec{x}} ],
\label{T8}
\end{equation}
where $F_{k\,\lambda}(\tau)$ is the (complex) mode function obeying Eq. (\ref{mot1A}) in the absence 
of anisotropic stress. The power spectrum introduced in Eq. (\ref{ST1}) become, in this specific case: 
\begin{eqnarray}
\langle 0| \hat{h}_{ij}(\vec{x},\tau) \, \hat{h}_{ij}(\vec{y},\tau)|0 \rangle &=& \int_{0}^{\infty} d\ln{k} \,{\mathcal P}_{\mathrm{T}}(k,\tau)\, \frac{\sin{kr}}{kr},\qquad r = |\vec{x} - \vec{y}|,
\label{T10a}\\
{\mathcal P}_{\mathrm{T}}(k,\tau) &=& \frac{4 \ell_{\mathrm{P}}^2\,\, k^3}{\pi^2} |F_{k}(\tau)|^2 =  4 \nu S_{h}(\nu,\tau),
\label{T10b}
\end{eqnarray}
where  $k = 2\pi \nu$ and $ S_{h}(\nu,\tau)$ is the so-called strain amplitude. The mode functions $F_{k}(\tau)$ are normalized 
during the inflationary phase \cite{max12} and the consistency relations of Eq. (\ref{r3}) can be easily derived in the case of 
conventional inflationary models by considering also the scalar fluctuations of the geometry.

When discussing the graviton spectra over various orders of magnitude in frequency it is more practical to deal with the spectral energy density of the relic gravitons in critical units\footnote{The natural logarithms will be denoted by $\ln$ while the common logarithms 
will be denoted by $\log$.} per logarithmic interval of wavenumber:
\begin{equation}
\Omega_{\mathrm{GW}}(k,\tau) = \frac{1}{\rho_{\mathrm{crit}}} \frac{d \rho_{\mathrm{GW}}}{d \ln{k}},\qquad 
\rho_{\mathrm{GW}} = \langle 0| T_{0}^{0}|0\rangle,
\label{T12}
\end{equation}
where $\rho_{\mathrm{crit}} = 3 H^2/\ell_{\mathrm{P}}^2$ is the critical 
energy density and $T_{\mu}^{\nu}$ denotes the energy-momentum pseudo-tensor of relic gravitons. 
The oscillations in $\Omega_{\mathrm{GW}}(k,\tau)$ for modes inside the Hubble radius 
are numerically less important and, as we shall remind later, it is possible 
to obtain rather accurate expressions for the transfer function across the matter-radiation transition.

The strain amplitude can be explicitly related to the spectral energy density in critical units 
and to the power spectrum. Using Eq. (\ref{T10b}), $\Omega_{\mathrm{GW}}(\nu,\tau)$ 
 can be expressed in terms of $S_{h}(\nu,\tau)$:
\begin{equation}
S_{h}(\nu,\tau) = \frac{3 {\mathcal H}^2}{4\pi^2 \nu^3} \Omega_{\mathrm{GW}}(\nu,\tau)  \to  7.981\times 10^{-43} \,\,\biggl(\frac{100\,\mathrm{Hz}}{\nu}\biggr)^3 \,\, h_{0}^2 \Omega_{\mathrm{GW}}(\nu,\tau_{0})\,\, \mathrm{Hz}^{-1},
\label{DEF2}
\end{equation}
where the second equality holds in the limit $\tau\to \tau_{0}$  where $\tau_{0}$ is the present value of the conformal time coordinate.
There is also a simple relation between the energy density of relic gravitons in critical units and the tensor power spectrum:
\begin{equation}
\Omega_{\mathrm{GW}}(k,\tau) = \frac{1}{\rho_{\mathrm{crit}}} \frac{d \rho_{\mathrm{GW}}}{d \ln{k}} = 
\frac{k^2}{12 H^2 a^2} {\mathcal P}_{\mathrm{T}}(k,\tau)\biggl[ 1 + {\mathcal O}\biggl(\frac{{\mathcal H}^2}{k^2}\biggr)\biggr].
\label{DEF1}
\end{equation}
The quantity in squared brackets at the right hand side of Eq. (\ref{DEF1}) is a consequence 
of the ambiguity inherent in all definitions of  the energy-momentum 
pseudo-tensor of the gravitons. Indeed, Eq. (\ref{DEF1}) holds at a generic conformal time and it includes all the modes that are inside 
the Hubble radius at the corresponding time; the correction ${\mathcal O}({\mathcal H}^2/k^2)$ depends on the specific 
definition of the energy-momentum pseudo-tensor of the relic gravitons. 

For instance, after getting rid of the tensor structure 
by making explicit the two physical polarizations,
Eq. (\ref{action1})  reduces to the action of two minimally coupled scalar fields in a conformally 
flat geometry of Friedmann-Robertson-Walker (FRW) type. As argued in Ref. \cite{ford}, the energy-momentum 
pseudo-tensor of relic gravitons in a FRW background can be formally obtained by taking the functional variation of Eq. (\ref{action1}) not with respect to the full metric but with respect to the background metric $\overline{g}_{\alpha\beta}$. In this case that the numerical factor in front of ${\mathcal H}^2/k^2$  in Eq. (\ref{DEF1}) is $1/2$. 
 In a complementary perspective \cite{BR}, the energy-momentum pseudo-tensor can be assigned by computing the second-order fluctuations of the Einstein tensor: this is exactly the Landau-Lifshitz approach to the energy-momentum pseudo-tensor of the gravitational waves 
 appropriately extended to curved backgrounds; the correction appearing in Eq. (\ref{DEF1}) is then given by $-7/2$ (instead of $1/2$).
The two  definitions seem very different but the energy densities and pressures derived within the two approaches give coincident 
 results as soon as the corresponding  wavelengths are inside the Hubble radius, i.e. $k > {\mathcal H}$.

\subsection{Frequency scales}

In addition to the pivot frequency $\nu_{p}$ of Eq. (\ref{fr1}) there are three other reference 
scales characterizing the cosmic background of relic gravitons: the frequency of matter-radiation equality (be it 
$\nu_{\mathrm{eq}}$), the frequency of neutrino decoupling (coinciding  approximately with the Hubble radius at the onset of big bang nucleosynthesis) and the maximal frequency of the spectrum $\nu_{\mathrm{max}}$ whose numerical value is fully determined, in the conventional setting, 
by the observed value of the tensor to scalar ratio.  

Denoting with $\Omega_{\mathrm{M}0}$ and $\Omega_{\mathrm{R}0}$ 
the present critical fraction of matter and radiation $\nu_{\mathrm{eq}}$ and $\nu_{\mathrm{bbn}}$ are:
\begin{eqnarray}
\nu_{\mathrm{eq}} &=& = 1.317 \times 10^{-17} \biggl(\frac{h_{0}^2 \Omega_{\mathrm{M}0}}{0.1364}\biggr) \biggl(\frac{h_{0}^2 \Omega_{\mathrm{R}0}}{4.15 \times 10^{-5}}\biggr)^{-1/2}\,\, \mathrm{Hz},
\label{EQ3}\\
\nu_{\mathrm{bbn}}&=& 
2.252\times 10^{-11} \biggl(\frac{g_{\rho}}{10.75}\biggr)^{1/4} \biggl(\frac{T_{\mathrm{bbn}}}{\,\,\mathrm{MeV}}\biggr) 
\biggl(\frac{h_{0}^2 \Omega_{\mathrm{R}0}}{4.15 \times 10^{-5}}\biggr)^{1/4}\,\,\mathrm{Hz}.
\label{EQ4}
\end{eqnarray}
where  $g_{\rho}$ denotes the effective number of relativistic degrees of freedom entering the total energy density of the plasma; by definition in the $\Lambda$CDM paradigm $\Omega_{\mathrm{M}0}$ is the sum of the baryon and of the CDM densities in critical units, i.e.
$\Omega_{\mathrm{M}0} =   \Omega_{\mathrm{c}0} + \Omega_{\mathrm{b}0}$.
The fiducial values appearing in Eqs. (\ref{EQ3}) and (\ref{EQ4}) are drawn from the best fit to the WMAP 9-yr data alone \cite{WMAP9} and within 
the $\Lambda$CDM paradigm (see also section \ref{sec3}).

The success of big-bang nucleosynthesis  
demands that, after neutrino decoupling, the Universe was already dominated by radiation. The standard $\Lambda$CDM paradigm  
makes stronger assumptions on the post-inflationary thermal history. Assuming, within the conventional lore, that 
the radiation dominates right at the end of inflation,  the maximal frequency of the cosmic background of relic gravitons  depends on the 
value of $r_{T}$:
\begin{equation}
\nu_{\mathrm{max}}  = 0.44 \,\biggl(\frac{r_{T}}{0.2}\biggr)^{1/4} 
\biggl(\frac{{\mathcal A}_{\mathcal R}}{2.41 \times 10^{-9}}\biggr)^{1/4}
\biggl(\frac{h_{0}^2 \Omega_{\mathrm{R}0}}{4.15 \times 10^{-5}}\biggr)^{1/4} \,\mathrm{GHz},
\label{EQ5}
\end{equation}
where ${\mathcal A}_{\mathcal R}$ has been introduced in Eq. (\ref{r2}).
The electroweak frequency $\nu_{\mathrm{ew}}$ and the TeV frequency that are,respectively, 
${\mathcal O}(10^{-3})$ Hz and ${\mathcal O}(10^{-6})$ Hz. More precisely we have that:
\begin{eqnarray}
\nu_{\mathrm{ew}} &=& 3.998\times 10^{-6} \biggl(\frac{g_{\rho}}{106.75}\biggr)^{1/4} \biggl(\frac{T_{*}}{100\,\,\mathrm{GeV}}\biggr) 
\biggl(\frac{h_{0}^2 \Omega_{\mathrm{R}0}}{4.15 \times 10^{-5}}\biggr)^{1/4}\,\,\mathrm{Hz},
\label{EW2}\\
\nu_{\mathrm{TeV}} &=& 4.819\times 10^{-3} \biggl(\frac{g_{\rho}}{228.75}\biggr)^{1/4} \biggl(\frac{T_{*}}{100\,\,\mathrm{TeV}}\biggr) 
\biggl(\frac{h_{0}^2 \Omega_{\mathrm{R}0}}{4.15 \times 10^{-5}}\biggr)^{1/4}\,\,\mathrm{Hz}.
\label{EW3}
\end{eqnarray}
In Eqs. (\ref{EW2}) and (\ref{EW3}) $g_{\rho}$ denote the fiducial values of the effective number of relativistic degrees of freedom
in the standard electroweak theory (i.e. $106.75$) and in the  minimal supersymmetric extension of the standard model (i.e. $228.75$).
In the conventional case $\nu_{\mathrm{ew}}$ and $\nu_{\mathrm{Tev}}$ do not play a special role but 
they become important whenever the high-frequency modifications of the spectra are concerned. 

The maximal frequency of the spectrum is closely related to the specific 
post-inflationary history that may not be exactly the one assumed in the $\Lambda$CDM case where 
after inflation the Universe reheats almost instantaneously. Different post-inflationary histories (see also section \ref{sec4}) change 
 the $N_{\mathrm{max}}$, i.e. maximal number of inflationary efolds accessible to large-scale CMB measurements \cite{LL}. 
The value of $N_{\mathrm{max}}$ can be derived by demanding that the inflationary event 
horizon redshifted at the present epoch coincides with the Hubble radius today:
\begin{equation}
e^{N_{\mathrm{max}}} = \frac{(2 \pi\, \Omega_{R 0} \,{\mathcal A}_{{\mathcal R}}\,r_{T})^{1/4}}{4}\, \biggl(\frac{M_{P}}{H_{0}}\biggr)^{1/2}\, \biggl(\frac{H}{H_{r}} \biggr)^{1/2- \gamma},
\label{eightha}
\end{equation}
where $\Omega_{R 0}$ is the present energy density of radiation in critical units, $H_{0}^{-1}$ is the Hubble radius today  and 
$\gamma$ controls the expansion rate during the post-inflationary phase. 

In terms of our fiducial set of parameters Eq. (\ref{eightha}) becomes:
 \begin{eqnarray}
 N_{\mathrm{max}} &=& 61.49 + \frac{1}{4} \ln{\biggl(\frac{h_{0}^2 \Omega_{R 0}}{4.15 \times 10^{-5}} \biggr)} - \ln{\biggl(\frac{h_{0}}{0.7}\biggr)}
 \nonumber\\
 &+& \frac{1}{4} \ln{\biggl(\frac{{\mathcal A}_{{\mathcal R}}}{2.41 \times 10^{-9}}\biggr)} + \frac{1}{4} \ln{\biggl(\frac{r_{T}}{0.2}\biggr)} + \biggl(\frac{1}{2} - \gamma\biggr) 
 \ln\biggl{(\frac{H}{H_{r}}\biggr)}.
 \label{eighthb}
\end{eqnarray} 
In Eqs. (\ref{eightha}) and (\ref{eighthb})  $\gamma$ accounts for the possibility of a delayed reheating terminating at a putative scale $H_{r}$
smaller than the Hubble rate during inflation. Since the reheating scale cannot be smaller than the one of nucleosynthesis, 
 $H_{r}$ can be as low as $10^{-44} M_{\mathrm{P}}$ (but not smaller) corresponding to a reheating scale occurring just prior to the formation of the light nuclei. If $\gamma - 1/2 >0$ (as it happens if $\gamma = 2/3$ when the post-inflationary background is dominated by dust), $N_{\mathrm{max}}$ diminishes in comparison with the sudden reheating (i.e. $H=H_{r}$) and $N_{\mathrm{max}}$ can become ${\mathcal O}(47)$. Conversely if $\gamma - 1/2 <0$ (as it happens in $\gamma = 1/3$ 
when the post-inflationary background is dominated by stiff sources), $N_{\mathrm{max}}$ increases. Finally, if $H_{r} = H$ (or, which is the same, if $\gamma=1/2$) there is a sudden transition 
between the inflationary regime and the post-inflationary epoch dominated by radiation. In spite of its dependence on ${\mathcal A}_{{\mathcal R}}$ and $r_{T}$, the value of $N_{\mathrm{max}}$ has then a theoretical error. Based on the previous considerations and on the maximal excursion of $\gamma$ we can say that $N_{\mathrm{max}} = 61.49 \pm  14.96$. The result of Eq. (\ref{EQ5}) has been derived in the case $\gamma=1/2$, i.e. in the 
sudden reheating approximation.

\renewcommand{\theequation}{3.\arabic{equation}}
\setcounter{equation}{0}
\section{Conventional inflationary spectra}
\label{sec3}
\subsection{Semianalytic considerations}
Between $\nu_{\mathrm{bbn}}$ and $\nu_{\mathrm{max}}$ there are roughly 20 orders of magnitude in frequency but 
 the cosmic graviton background,  in the concordance scenario, is solely determined by $r_{T}$. This statement 
  is approximately true if we neglect neutrino free streaming and other 
 comparable sources of further damping. Even before actual formulation of inflationary models, the pioneering works of Grishchuk \cite{gr} 
 have	shown	that, under certain conditions, gravitational-wave amplification can occur in an expanding universe and can lead to observable effects today. 
Since then various analyses of the relic graviton backgrounds have been reported in the literature within different perspectives (see e.g. 
\cite{rub,FS1,paul,TR1,TR2,TR3,max12} for an incomplete list of references). 
Analytical approximations are somehow simpler but not as accurate. We shall therefore employ the numerical methods 
described in \cite{max12} that seem the more suitable for the present ends. 

It was noticed in Ref. \cite{max12} that it is preferable to compute 
directly the transfer function of spectral energy density $\Omega_{\mathrm{GW}}(\nu, \tau)$ rather 
than computing the transfer function for the power spectrum ${\mathcal P}_{T}(\nu,\tau)$ and then use it 
to estimate $\Omega_{\mathrm{GW}}(\nu,\tau)$ from Eq. (\ref{DEF1}) and its descendant. 
To exemplify this technique we neglect all the potential sources of further damping and just compute the low-frequency tail of the spectrum. 

Since around equality the expansion rate changes, the spectral energy density will have a break in its slope so that $\Omega_{\mathrm{GW}}(\nu,\tau)$ will ultimately 
depend on $\nu_{\mathrm{eq}}$. In addition, recalling Eq. (\ref{EQ5}), the spectrum will be exponentially damped for 
typical frequencies larger than $\nu_{\mathrm{max}}$. The semi-analytic result for the spectral energy density in critical units is: 
\begin{eqnarray}
h_{0}^2 \Omega_{\mathrm{GW}}(\nu,\tau_{0}) &=& {\mathcal N}_{\rho}  \,T^2_{\rho}(\nu/\nu_{\mathrm{eq}})\, r_{\mathrm{T}}\, \biggl(\frac{\nu}{\nu_{\mathrm{p}}}\biggr)^{n_{\mathrm{T}}} e^{- 2 \beta \frac{\nu}{\nu_{\mathrm{max}}}}, 
\nonumber\\
{\mathcal N}_{\rho} &=& 4.165 \times 10^{-15} \biggl(\frac{h_{0}^2 \Omega_{\mathrm{R}0}}{4.15\times 10^{-5}}\biggr).
\label{EQ20}
\end{eqnarray}
The parameter $\beta= {\mathcal O}(1)$ appearing in Eq. (\ref{EQ5}) depends upon the width of the transition between the inflationary phase and the subsequent radiation dominated phase. Numerically, for different widths of the smooth transitions between we can estimate $0.5 <\beta < 6.33$ \cite{max12}.
In Eq. (\ref{EQ20}) the transfer function across equality is given by:
\begin{equation}
T_{\rho}(\nu/\nu_{\mathrm{eq}}) = \sqrt{1 + c_{1}\biggl(\frac{\nu_{\mathrm{eq}}}{\nu}\biggr) + b_{1}\biggl(\frac{\nu_{\mathrm{eq}}}{\nu}\biggr)^2},\qquad c_{1}= 0.5238,\qquad
b_{1}=0.3537.
\label{EQ19}
\end{equation}
Equation (\ref{EQ19}) is obtained by integrating Eq. (\ref{mot1A}) across the radiation-matter transition and by computing 
$\Omega_{GW}(\nu,\tau)$ for different frequencies. The initial conditions for the mode functions are the ones 
obtained, for the corresponding frequencies, from the solutions of Eq. (\ref{mot1A}) during inflation and in the absence of anisotropic 
stress.

Using Eq. (\ref{EQ20}) into Eqs. (\ref{DEF1}) and (\ref{DEF2}) 
 it is immediate to compute the power spectrum ${\mathcal P}_{T}(\nu,\tau_{0})$ or the strain amplitude $S_{h}(\nu,\tau_{0})$.
As we expect from the standard analytic estimates (see e.g. \cite{mg1,mg2}) $T^2_{\rho} \to 1$ for $\nu \gg \nu_{\mathrm{eq}}$ while
$T^2_{\rho} \to (\nu/\nu_{\mathrm{eq}})^{-2}$ for $\nu \ll \nu_{\mathrm{eq}}$.  Note that the limit $\nu \ll  \nu_{\mathrm{eq}}$
is not completely physical since, in the realistic situation, $\nu_{p}$ and $\nu_{\mathrm{eq}}$ are different 
but can be numerically close depending on the choice of the pivot scale.

\subsection{Normalization of the spectra}
Equation (\ref{EQ20}) shows that the absolute normalization of the spectra is 
reduced to the specific value of $r_{T}$. In the realistic situation, however, the spectra depend on other 
late time parameters that are anyway fixed in the concordance paradigm.
The WMAP 9-yr data \cite{WMAP9} alone analyzed in the 
light of the $\Lambda$CDM scenario (supplemented by tensors) imply $r_{\mathrm{T}} <0.38$;
this determination  is consistent with the Bicep2 data. In this case the remaining cosmological parameters are:
\begin{equation}
( \Omega_{\mathrm{b}0}, \, \Omega_{\mathrm{c}0}, \Omega_{\mathrm{de}0},\, h_{0},\,n_{\mathrm{s}},\, \epsilon_{\mathrm{re}}) \equiv 
(0.0442,\, 0.210,\, 0.746,\,0.726,\, 0.992,\,0.091),
\label{par1}
\end{equation}
while ${\mathcal A}_{\mathcal R} = 2.26\times 10^{-9}$. If we take the WMAP 9-yr data alone with no tensors we shall have instead:
\begin{equation}
( \Omega_{\mathrm{b}0}, \, \Omega_{\mathrm{c}0}, \Omega_{\mathrm{de}0},\, h_{0},\,n_{\mathrm{s}},\, \epsilon_{\mathrm{re}}) \equiv 
(0.0463,\, 0.233,\, 0.721,\,0.7,\, 0.972,\,0.089),
\label{par2}
\end{equation}
with ${\mathcal A}_{\mathcal R} = 2.41\times 10^{-9}$. The numerical values of the parameters estimated within the Planck data are:
\begin{equation}
( \Omega_{\mathrm{b}0}, \, \Omega_{\mathrm{c}0}, \Omega_{\mathrm{de}0},\, h_{0},\,n_{\mathrm{s}},\, \epsilon_{\mathrm{re}}) \equiv 
(0.0490,\, 0.2693,\, 0.6817,\,0.6704,\, 0.9619,\,0.089).
\label{par3}
\end{equation}
The Planck collaboration could not determine the cosmological parameters 
without assuming, either directly or indirectly, the polarization measurements and the WMAP 9-yr data. We shall therefore 
adopt Eq. (\ref{par2}) as fiducial set of cosmological parameters. Slightly different determinations of the pivotal parameters have a minor impact on the normalization of the cosmic graviton background, as we shall see.  

Let us now  discuss the value of $r_{T}$ as a function of the (conventional) pivot scale.
It is desirable, when comparing different measurements, that 
each experiment chooses consistently a pivot scale to present its own data. If a certain experiment 
measures $r_{T}$ at a given pivot scale $\overline{k}_{p}$ and if the tensor and the scalar power spectra are both assigned 
at the {\em same} pivot scale, then it is obvious that $\overline{r}_{T}(\overline{k}_{p})$ does not depend on the scalar 
and tensor spectral indices:
\begin{equation}
\overline{r}_{T}(\overline{k}_{p})  = \frac{\overline{{\mathcal A}}_{T}}{\overline{{\mathcal A}}_{{\mathcal R}}},
\end{equation}
where, within the notations established in Eq. (\ref{r2}),  $\overline{{\mathcal P}}_{{\mathcal R}}(k) = \overline{{\mathcal A}}_{{\mathcal R}} (k/\overline{k}_{p})^{n_{s} -1}$ and $\overline{{\mathcal P}}_{T}(k) = \overline{{\mathcal A}}_{T} (k/\overline{k}_{p})^{n_{T}}$.
If we now wish to determine the tensor to scalar ratio at a different pivot scale we must use the generic definition 
of $r_{T}(k)$ at an arbitrary scale. The resulting expression will both depend on $n_{s}$ and $n_{T}$:
\begin{equation}
r_{T}(k,\,k_{p},\,n_{s},\,n_{T}) = \frac{{\mathcal A}_{T}}{{\mathcal A}_{{\mathcal R}}} \biggl(\frac{k}{k_{p}}\biggr)^{1 + n_{T} - n_{s}},
\label{rr2}
\end{equation}
where ${\mathcal A}_{T}$ and ${\mathcal A}_{{\mathcal R}}$ are the amplitudes the tensor and scalar power spectra at $k_{p}$ (rather 
then at $\overline{k}_{p}$).  The tensor to scalar ratio at $k_{p}$ (i.e. $r_{T}$) will then be related to the tensor to scalar ratio at $\overline{k}_{p}$ (i.e. $\overline{r}_{T}$) as: 
\begin{equation}
r_{T} = \overline{r}_{T} \biggl(\frac{k_{p}}{\overline{k}_{p}}\biggr)^{1 + n_{T} - n_{s}}.
\label{rr3}
\end{equation}
In the realistic case we could fix, for instance, $\overline{k}_{p} = 0.05\, \mathrm{Mpc}^{-1}$ (as assumed by Bicep2)
while $k_{p} =0.002\, \mathrm{Mpc}^{-1}$ as assumed by Planck. This choice implies that 
 $\overline{r}_{T} =0.2$ and $n_{T} = - \overline{r}_{T}/8$ (from the consistency relations). To determine $r_{T}$ at $k_{p}$ 
 we still need the value of $n_{s}$ that we can take from Planck (i.e. $n_{s} = 0.961$). 
 Applying Eq. (\ref{rr3}) we shall have that $r_{T} = 0.19$ at $k_{p}$.  
 
Occasionally the same experiment assigns the scalar and the tensor power 
spectra at different pivot scales.  For instance the Planck 
experiment assigns the scalar power spectrum at $\overline{k}_{p} =0.05 \, \mathrm{Mpc}^{-1}$  and the 
tensor to scalar ratio $r_{T}$ at $k_{p} =0.002\, \mathrm{Mpc}^{-1}$. These choices are possible but 
become contrived when the results of different experiments have to be compared. It is finally unclear if
the Bicep2 collaboration assumes the consistency relations or simply sets $n_{T} =0$ in the analysis. 
For all these reasons (and since the issues on the proper subtraction of the foregrounds are not yet clearly settled) we 
shall assume, in the present calculation, $r_{T} =0.2$ with a potential indetermination leading to slightly lower 
values at $r_{T}$. This indetermination would be anyway hardly visible in the forthcoming plots.

\subsection{Concordance paradigm}
\begin{figure}[!ht]
\begin{center}
      \epsfxsize = 11 cm  \epsffile{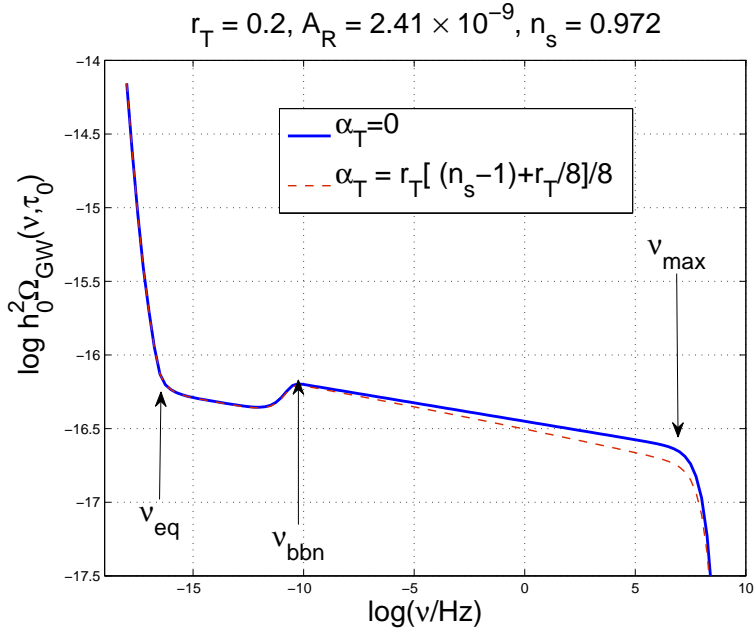}
\end{center}
\caption[a]{The cosmic background of relic gravitons is illustrated as a function of the comoving frequency. Common logarithms are used on the vertical and on the horizontal axis. The full line denotes the case where the consistency relations are enforced and the tensor 
spectral index does not run. The dashed line illustrates the case $\alpha_{T} \neq 0$. The fiducial set of parameters has been fixed as in Eq. (\ref{par2}).}
\label{Figure1}
\end{figure}

In Fig. \ref{Figure1} the normalized cosmic background of relic gravitons is illustrated for the fiducial value of cosmological parameters 
of Eq. (\ref{par2}). The typical frequencies introduced in Eqs. (\ref{EQ3}), (\ref{EQ4}) and (\ref{EQ5}) have been indicated 
for convenience.
The full line in Fig. \ref{Figure1} denotes the cosmic graviton background determined by enforcing the consistency relations of Eq. (\ref{r3}) and in the absence of any running of the tensor spectral index (i.e. $\alpha_{T} =0$ in Eq. (\ref{break3})). 
Always in Fig. \ref{Figure1}, the barely visible dashed line denotes the case where the spectral index runs according to Eq. (\ref{break3}). 
The estimate of Eq. (\ref{EQ20}) gives a value of $h_{0}^2 \Omega_{\mathrm{GW}}(\nu,\tau)$ that is significantly larger than the one implied by the numerical result, as it should be clear by looking at Fig. \ref{Figure1}.  The reason for this mismatch between the two estimates is due to three physical effects that are included in Fig. \ref{Figure1} but that 
are absent from Eq. (\ref{EQ20}). 

The shallow suppression of the spectrum for $\nu< \nu_{\mathrm{bbn}}$ is due to the neutrino free 
streaming. The neutrinos free stream, after their decoupling, and the effective energy-momentum tensor acquires, to first-order in the amplitude 
of the plasma fluctuations, an anisotropic stress, $\Pi_{ij}$ that has been already included for illustration in Eq. (\ref{mot1A}). 
This effect leads to an integro-differential equation which has been specifically analyzed, for instance, in 
\cite{wnu1,wnu2}. The overall effect of collisionless particles is a reduction 
of the spectral energy density of the relic gravitons. Assuming that the only collisionless 
species in the thermal history of the Universe are the neutrinos, the amount 
of suppression can be parametrized by the function
\begin{equation}
{\mathcal F}(R_{\nu}) = 1 -0.539 R_{\nu} + 0.134 R_{\nu}^2,
\label{ANIS3}
\end{equation}
where $R_{\nu}$ is the fraction of neutrinos in the radiation plasma, i.e. 
\begin{equation}
R_{\nu} = \frac{r}{r + 1}, \qquad r = 0.681 \biggl(\frac{N_{\nu}}{3}\biggr),\qquad R_{\gamma} + R_{\nu} = 1. 
\label{ANIS4}
\end{equation}
In the case $R_{\nu}=0$ (i.e. in the absence of collisionless particles) there is no suppression. If, on the contrary, 
$R_{\nu} \neq 0$ the suppression can even reach one order of magnitude. In the case $N_{\nu} = 3$, 
$R_{\nu} = 0.405$ and the suppression of the spectral energy density is proportional 
to ${\mathcal F}^2(0.405)= 0. 645$. This suppression will be effective for relatively 
small frequencies which are larger than $\nu_{\mathrm{eq}}$ and smaller than $\nu_{\mathrm{bbn}}$.

The second effect included in Fig. \ref{Figure1} is the damping effect associated 
with the (present) dominance of the dark energy component. The redshift of $\Lambda$-dominance is given by 
\begin{equation}
1 + z_{\Lambda} = \biggl(\frac{a_{0}}{a_{\Lambda}}\biggr) = 
 \biggl(\frac{\Omega_{\mathrm{de}}}{\Omega_{\mathrm{M}0}}\biggr)^{1/3},
\label{LAM1}
\end{equation}
where, in the concordance paradigm, $\Omega_{\mathrm{de}} \equiv \Omega_{\Lambda}$. 
In principle there should be a breaking in the spectrum for the modes reentering the Hubble radius after $\tau_{\Lambda}$  (i.e. $k< k_{\Lambda} = H_{\Lambda} a_{\Lambda}$). Since for $\tau > \tau_{\Lambda}$ the Hubble rate is constant in the $\Lambda$CDM case
Eq. (\ref{LAM1}) implies that $k_{\Lambda} = (\Omega_{\mathrm{M}0}/\Omega_{\Lambda})^{1/3}k_{\mathrm{H}}$ 
where $k_{\mathrm{H}} = a_{0} H_{0}$. The explicit value of $\nu_{\Lambda}$ is 
\begin{equation}
 \nu_{\Lambda} = 2.58 \times 10^{-19}  \biggl(\frac{h_{0}}{0.7}\biggr) \biggl(\frac{\Omega_{\mathrm{M}0}}{0.2793}\biggr)^{1/3} \biggl(\frac{\Omega_{\Lambda}}{0.721}\biggr)^{1/3} \,\, \mathrm{Hz}.
\label{LAM3}
\end{equation}
Since the frequency interval between $\nu_{\mathrm{H}}$ and $\nu_{\Lambda}$ is tiny the modification to the slope is practically irrelevant. However,  the adiabatic damping of the tensor mode function across $\tau_{\Lambda}$ reduces the amplitude of the spectral energy density by a factor $(\Omega_{\mathrm{M}0}/\Omega_{\Lambda})^2$. For the fiducial choice of parameters of Eqs. (\ref{par2}) and (\ref{LAM3}) we have that 
the suppression is of the order of $0.10$. This figure is comparable with the suppression due to 
the neutrino free streaming. These effects have been discussed in \cite{max12} (see also \cite{TR1}).  

There is also a third effect reducing the quasi-flat plateau of Fig. \ref{Figure1} and it has to do with the variation of the effective number of relativistic species. Recall, indeed, that 
the total energy density and the total entropy density of the plasma can be written as 
\begin{equation}
\rho_{\mathrm{t}} = g_{\rho}(T) \frac{\pi^2}{30} T^4,\qquad s_{\mathrm{t}} = g_{\mathrm{s}}(T) \frac{2 \pi^2}{45} T^3.
\label{EFF1}
\end{equation}
For temperatures much larger than the top quark mass, all the known species of the minimal standard model of particle interactions are in local thermal equilibrium, then $g_{\rho} = g_{\mathrm{s}} = 106.75$. Below, $T \simeq 175$ GeV the various species 
start decoupling,  and the 
time evolution of the number of relativistic degrees of freedom effectively changes the evolution of the Hubble rate. 
In principle if a given mode $k$ reenters the Hubble radius at a temperature $T_{k}$ the spectral energy density 
of the relic gravitons is (kinematically) suppressed by a factor which can be written as 
 \begin{equation}
 \biggl(\frac{g_{\rho}(T_{k})}{g_{\rho0}}\biggr)\biggl(\frac{g_{\mathrm{s}}(T_{k})}{g_{\mathrm{s}0}}\biggr)^{-4/3}.
 \label{EFF2}
 \end{equation}
At the present time  $g_{\rho0}= 3.36$ and $g_{\mathrm{s}0}= 3.90$. In general terms the effect parametrized by 
Eq. (\ref{EFF2}) will cause a frequency-dependent suppression, i.e. a further modulation of the spectral 
energy density $\Omega_{\mathrm{GW}}(\nu,\tau_{0})$.  The maximal suppression one can expect 
can be obtained by inserting into Eq. (\ref{EFF2}) the highest possible number of degrees of freedom. 
So, in the case of the minimal standard model this would imply that the suppression (on $\Omega_{\mathrm{GW}}(\nu,\tau_{0})$)
 will be of the order of $0.38$. In popular supersymmetric extensions of the minimal standard models $g_{\rho}$ and $g_{s}$  can be as high as, approximately, $230$. This will bring down the figure given above to $0.29$.
 
\begin{figure}[!ht]
\begin{center}
      \epsfxsize = 11 cm  \epsffile{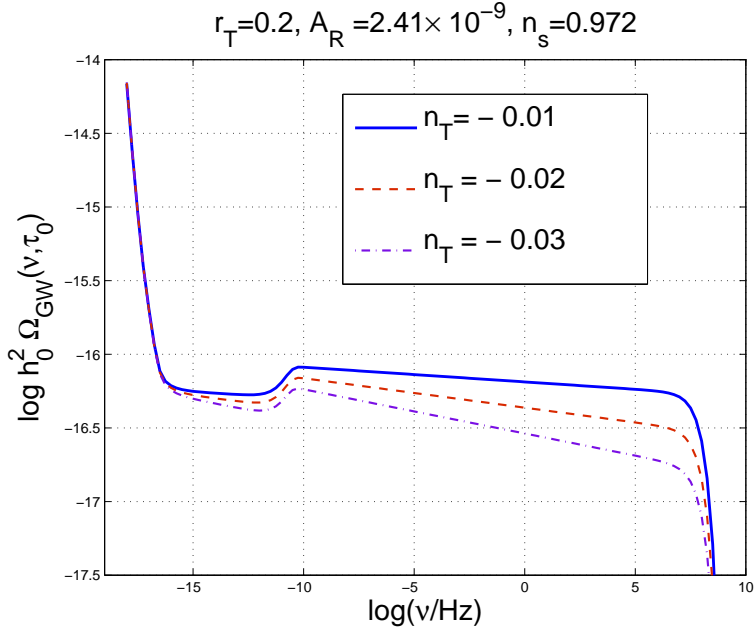}
\end{center}
\caption[a]{The cosmic graviton background in the case when the 
spectral indices are assigned independently from the value of the tensor to scalar ratio. 
The common logarithm is reported on both axes. The fiducial set of cosmological parameters 
is the one given in Eq. (\ref{par2}), as in the case of Fig. \ref{Figure1}.}
\label{Figure2}
\end{figure}

All the effects discussed above in this section have been conservatively included by avoiding, for instance, an artificial 
largeness of the effective number of relativistic degrees of freedom. We just assumed the particle content of the electroweak
standard model. All the neutrino species have been taken to be massless since this is what we posit in the 
$\Lambda$CDM scenario.  Deviations from these assumptions imply further reductions of the plateau of Fig. \ref{Figure1}. 
In Fig. \ref{Figure2} we illustrate the cosmic graviton background by assigning independently the tensor spectral index $n_{T}$ and the 
tensor to scalar ratio $r_{T}$. This is the simplest way to break explicitly the consistency relations of Eq. (\ref{r3}). The results 
of Figs. \ref{Figure1} and \ref{Figure2} are clearly consistent but quantitatively different. 
The violation of the consistency relations can be justified in the context of the 
protoinflationary dynamics where the initial conditions of the scalar and tensor inhomogeneities 
of the geometry may be slightly asymmetric. An example along this direction is given by a 
thermal state of fluid phonons and gravitons \cite{vio}.
 
The results of the present section can be used for different purposes. They can be compared, for instance, with the sensitivity of wide-band interferometers to a quasi-flat spectral energy density. 
The specific frequency at which $\Omega_{\mathrm{GW}}(\nu,\tau_{0})$ is computed is $\nu_{\mathrm{LV}}= 100$Hz.
The subscript LV is a shorthand notation for LIGO/Virgo. In the case of an exactly scale invariant spectrum
the correlation of the two (coaligned) LIGO detectors with 
central corner stations in Livingston (Lousiana) and in Hanford 
(Washington) might reach a sensitivity given by  \cite{corr}
 \begin{equation}
h_0^2\,\, \Omega_{\mathrm{GW}}(\nu_{\mathrm{LV}},\tau_{0}) \simeq 6.5 \times 10^{-11} \,\, 
\biggl(\frac{1\,\,\mathrm{yr}}{T} \biggr)^{1/2}\,\,\mathrm{SNR}^2, \qquad \nu_{\mathrm{LV}} =0.1 \,\, \mathrm{kHz}
\label{SENS}
\end{equation} 
where $T$ denotes the observation time and $\mathrm{SNR}$ is the signal to noise ratio.  Equation (\ref{SENS}) is in close agreement with the 
sensitivity of the advanced LIGO apparatus to an exactly scale-invariant spectral energy density \cite{LIGOS1,LIGOS2,LIGOS3} 
(see also \cite{virgoligo,LIGO4}). Equation (\ref{SENS}) must be compared with the values obtainable from the fiducial choice of parameters 
illustrated in Fig. \ref{Figure1}; in the absence of running we shall have 
\begin{equation}
  h_0^2\,\, \Omega_{\mathrm{GW}}(\nu_{\mathrm{LV}},\tau_{0}) = 3.15 \times 10^{-17}.
\label{SENS1}
\end{equation}
If the running is added the figure of Eq. (\ref{SENS1}) decreases to $2.71\times10^{-17}$. Finally, if the suppression 
associated with the dark energy dominance is neglected the value of the spectral energy density increases even further to $2.10\times 10^{-16}$.
We stress that these figures are obtained by assuming the fiducial set of parameters implied by the WMAP 9-yr data. Slightly 
different determinations of the parameters do not lead to appreciable differences.
It is important to stress that the results of Figs. \ref{Figure1} and \ref{Figure2} or even the estimate of Eq. (\ref{SENS1}) 
are qualitatively different from the well known analytic approximations that are still used by the experimental collaborations 
for their sensitivity goals (see e.g. \cite{LIGO4}). Now the low-frequency bounds should be replaced by the Bicep2 determination 
of the tensor to scalar ratio. In this sense Eq. (\ref{SENS1}) is a consequence of the concordance model and not just 
an estimate of the largest theoretical signal compatible with the low-frequency bounds.

\renewcommand{\theequation}{4.\arabic{equation}}
\setcounter{equation}{0}
\section{Uncertainties at high frequencies}
\label{sec4}
The results of the preceding section suggest that the relic graviton 
background predicted by the concordance paradigm is unobservable 
by operating wide-band interferometers.
The minuteness of $h_{0}^2 \Omega_{\mathrm{GW}}(\nu_{\mathrm{LV}},\tau_{0})$ stems directly from the 
assumption that the inflationary phase is  suddenly followed by the radiation-dominated phase. This result 
is already relevant if we ought to calibrate the sensitivities of future instruments, at low or high
frequencies. In this section we shall discuss the theoretical uncertainties arising in the high-frequency 
tail of the cosmic background of relic gravitons.
\begin{figure}[!ht]
\begin{center}
      \epsfxsize = 11 cm  \epsffile{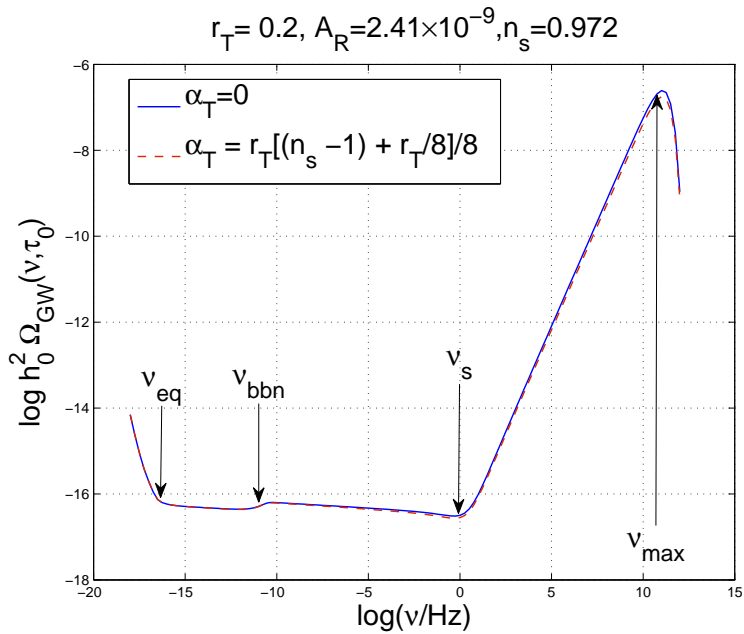}
\end{center}
\caption[a]{The cosmic graviton background in the case of a post-inflationary phase 
stiffer than radiation. The full line illustrates the case when the 
tensor spectral index does not run at low frequencies.
The common logarithm is reported on both axes. The fiducial values of the cosmological 
parameters have been chosen as in Eq. (\ref{par2}).}
\label{Figure3}
\end{figure}

Let us first consider the known possibility that the inflationary epoch is not suddenly 
followed by a  radiation-dominated phase. In Fig. \ref{Figure3} we illustrate 
the case where the transition from inflation to radiation is not sudden but rather delayed by a 
long post-inflationary phase dominated by a stiff fluid with sound speed coinciding 
with the speed of light \cite{mg1} (see also \cite{zel1,pee}). In this case the total barotropic 
index during the stiff phase is $w_{t} = 1$.

The notations of Fig. \ref{Figure3} are the same of Fig. \ref{Figure2}: the full line denotes 
the case without running of the spectral index while the barely visible dashed curve includes the 
effect of the running. The results of Fig. \ref{Figure3} have been obtained by imposing the Bicep2 normalization 
and by integrating directly the evolution equations in the case of a modified 
thermal history including the late-time domination of dark energy, the effect 
of neutrino free streaming and the evolution of the relativistic degrees of freedom of the 
plasma discussed in section \ref{sec3}. 

\subsection{Stiff phases in the early Universe}
The perspective adopted here is rather heuristic: stiff phases (as well 
as waterfall transitions) are just possible examples of high-frequency uncertainties 
in the cosmic graviton background.  In spite of the previous statement, 
the motivation for the existence of a stiff phase in the early stages 
of the evolution of the Universe is twofold. We shall account hereunder of both 
perspectives. 
    
   The first motivation for the existence of a stiff phase is indirect and it stems from 
    the current ignorance on the thermodynamic history of the Universe 
    for temperatures larger than the MeV and this is, somehow, the 
    perspective adopted by Zeldovich \cite{zel1} (see also first paper in Ref. \cite{mg1} second paper in Ref. \cite{gr}). 
    It is not implausible that prior to the epoch of radiation dominance there was a phase expanding at a rate 
    slower than radiation. The slowest possible rate of expansion 
    occurs when the sound speed of the medium coincides with the 
    speed of light. Expansion rates even slower than the ones of the stiff phase can only be realized 
     when the sound speed exceeds the speed of light. 
     This possibility is however not compatible with the standard notion of causality.
    The plausible range for the existence of such a phase 
    is between the end of inflation and the formation of the light nuclei.
    If the dominance of radiation is to take place already by the 
    time of formation of the baryon asymmetry, then the  
    onset of radiation dominance increases from few MeV to the TeV range. 
    
    The second motivation for the existence of a stiff phase stems 
    from a detailed consideration of specific models leading 
    to the late dominance of an effective cosmological term.
    In quintessence scenarios the present dominance of a cosmological
    term is translated into the late-time dominance of the potential of a 
    scalar degree of freedom that is called quintessence (see e.g. \cite{wein}). 
    If we also demand the existence of an early inflationary phase to account, among other things, 
     for  the existence of large-scale inhomogeneities, we are in the situation when the 
    inflaton potential did dominate at early times while the quintessence potential does dominate 
    at late times (see second and third papers in Ref. \cite{mg1} and \cite{pee}). In between 
    the scalar kinetic term of inflaton/quintessence field dominates the background.
    When the inflaton and the quintessence field are identified
    the existence of this phase is explicitly realized \cite{pee}, in other models it can be anyway 
    speculated.
        
    The slope of the stiff phase can also be parametrized in terms of the barotropic index 
    $w_{t}$ that does not need to coincide with $1$ (implying that the sound speed and the 
    speed of light coincide exactly). For instance we can imagine that $ 1/3 < w_{t} \leq 1$ \cite{max12}
    implying that the effective fluid driving the background geometry is stiffer than radiation 
    without being driven by the kinetic energy of the inflaton/quintessence field.

\subsection{Some specific examples}
The model illustrated in Fig. \ref{Figure3} involves, on top of $r_{T}$, 
only two supplementary parameters that can be identified with the the frequency $\nu_{\mathrm{s}}$ and with 
the slope of $\Omega_{\mathrm{GW}}(\nu,\tau)$ for $\nu> \nu_{\mathrm{s}}$. In this sense 
the case illustrated in Fig. \ref{Figure3} is next to minimal.
The value of $\nu_{\mathrm{s}}$  marks the border between the region of intermediate frequencies and the high-frequency tail of the spectrum 
in the same way as $\nu_{\mathrm{eq}}$ defines the range of the low-frequency branch.
The transfer function for the spectral energy density is given by \cite{max12}
\begin{equation}
T_{\rho}^2(\nu/\nu_{\mathrm{s}}) = 1.0  + 0.204\,\biggl(\frac{\nu}{\nu_{\mathrm{s}}}\biggr)^{1/4} - 0.980 \,\biggl(\frac{\nu}{\nu_{\mathrm{s}}}\biggr)^{1/2}  + 3.389 \biggl(\frac{\nu}{\nu_{\mathrm{s}}}\biggr) -0.067\,\biggl(\frac{\nu}{\nu_{\mathrm{s}}}\biggr)\ln^2{(\nu/\nu_{\mathrm{s}})},
\label{TT}
\end{equation}
and it can be used to derive semianalytic estimates in analogy with what has been discussed in Eq. (\ref{EQ20}). 
In specific models (see e.g. \cite{mg1}) the  frequencies $\nu_{\mathrm{s}}$ and $\nu_{\mathrm{max}}$ depend on a single parameter $Q$:
\begin{eqnarray}
\nu_{\mathrm{max}} &=& 1.177 \times 10^{11} \, Q^{-1} 
\biggl(\frac{h_{0}^2 \Omega_{\mathrm{R}0}}{4.15 \times 10^{-5}}\biggr)^{1/4}\,\,\mathrm{Hz},
\label{FFmax}\\
\nu_{\mathrm{s}} &=& 11.10 \,\,Q^3 \,\,\biggl(\frac{r_{T}}{0.2}\biggr) \,\biggl(\frac{{\mathcal A}_{\mathcal R}}{2.41 \times 10^{-9}}\biggr) \,\biggl(\frac{h_{0}^2 \Omega_{\mathrm{R}0}}{4.15 \times 10^{-5}}\biggr)^{1/4}\,\,\mathrm{Hz}.
\label{FFs}
\end{eqnarray}
The natural choices of the Q parameter are determined by the need of either 
    preserving the light nuclei or by the need of preserving the baryon asymmetry. In the 
    first case the turnaround frequency can be as small as $10^{-9}$ Hz. In the second case it is 
    of the order of few mHz ($1$ mH= $10^{-3}$ Hz). These considerations 
    are model independent and hold for any high-frequency modification of the 
    graviton spectrum. Thus if the parameter $Q$ is taken as a free parameter we must always demand that $\nu_{\mathrm{s}} > \nu_{\mathrm{bbn}}$ and,
in some cases, it is necessary to impose further conditions such as 
$\nu_{\mathrm{s}}> \nu_{\mathrm{ew}}$ or even  $\nu_{\mathrm{s}}>  \nu_{\mathrm{TeV}}$.
 
\begin{figure}[!ht]
\begin{center}
      \epsfxsize = 11 cm  \epsffile{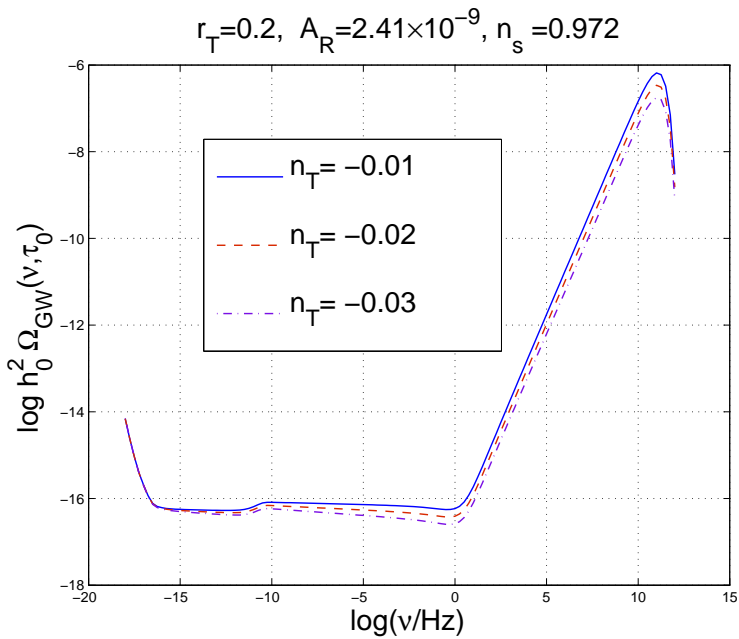}
\end{center}
\caption[a]{The cosmic graviton background in the case of a stiff post-inflationary phase. The low-frequency 
slope of the spectra are assigned independently of the value of $r_{T}$. The common logarithm is used on both axes. }
\label{Figure4}
\end{figure}

   In the context of quintessential inflationary models 
    we can determine Q on a theoretical basis by assuming that the dominance 
    of radiation is triggered by the back-reaction of some massless fields present 
    at the end of inflation. In explicit models and without fine tunings $Q = {\mathcal O}(1)$ but a bit smaller than $1$. In the case of \cite{mg1}, for instance $Q=0.37$ so that $\nu_{\mathrm{s}} = {\mathcal O}(4.3)$ Hz and this is the case 
reported in Fig. \ref{Figure3}. In Fig. \ref{Figure4} the high-frequency uncertainties are illustrated in the case when the tensor spectral index $n_{T}$ and the tensor to scalar ratio are independently assigned. 
The value $Q= 0.37$ (implying $\nu_{\mathrm{s}} = {\mathcal O}(4.3)$ Hz ) arises by assuming that  
 $N_{\mathrm{eff}}$ nearly massless degrees of freedom will be amplified with typical energy density $H^4$ scaling as $a^{-4}$. But the energy density 
 of the background will scale, in the case of Fig. \ref{Figure3}, as $a^{-6}$. The value of $Q$ will then be determined in this case 
 as $(N_{\mathrm{eft}}/(480 \pi^2))^{1/4}$. For $N_{\mathrm{eff}} = 90$ we will have $Q= 0.37$; for $N_{\mathrm{eff}} = 106.75$ (corresponding to the relativistic degrees of freedom in the minimal standard model)
 we will have $Q= 0.38$; if $N_{\mathrm{eff}}= 10^{3}$ we will then have $Q= 0.6$. This is why we said that if we do not tune $N_{\mathrm{eff}}$ to be 
 much larger than ${\mathcal O}(100)$ the turnaround frequency will be (in this explicit case) ${\mathcal O}(\mathrm{Hz})$ that turns out to be 
 automatically larger than  $\nu_{\mathrm{ew}}$; in this case the plasma will be already dominated by radiation at the approximate time 
 of the electroweak phase transition.
 
 \subsection{Model-independent considerations}
To appreciate the relevance of a benchmark value for $r_{T}$, 
it is interesting to compare the strategy leading to Fig. \ref{Figure3} with the conventional 
way of dealing with growing spectra of cosmic gravitons before the Bicep2 
observation. When the value of $r_{T}$ was unknown (and potentially 
very small) the models leading to a growing spectral energy density were customarily normalized
at high frequencies by imposing, simultaneously, the bounds stemming from the 
pulsar timing measurements and from the number of massless species 
at big-bang nucleosynthesis. The pulsar timing constraint demands \cite{pulsar1,pulsar2} 
\begin{equation}
\Omega(\nu_{\mathrm{pulsar}},\tau_{0}) < 1.9\times10^{-8},\qquad 
\nu_{\mathrm{pulsar}} \simeq \,10^{-8}\,\mathrm{Hz},
\label{PUL}
\end{equation}
where $\nu_{\mathrm{pulsar}}$ roughly corresponds to the inverse 
of the observation time along which the pulsars timing has been monitored. Such a bound is clearly not constraining for 
Fig. \ref{Figure3}. 
A similar conclusion can be drawn in the case of the big-bang nucleosynthesis constraint stipulating 
that the bound on the extra-relativistic species at the time of big-bang nucleosynthesis can be 
translated into a bound on the cosmic graviton backgrounds \cite{bbn1}. For historical reasons 
this constraint is often expressed in terms of $\Delta N_{\nu}$ representing the contribution of supplementary 
neutrino species but the extra-relativistic species do not need to be fermionic. If the additional species are 
relic gravitons we have: 
\begin{equation}
h_{0}^2  \int_{\nu_{\mathrm{bbn}}}^{\nu_{\mathrm{max}}}
  \Omega_{{\rm GW}}(\nu,\tau_{0}) d\ln{\nu} = 5.61 \times 10^{-6} \Delta N_{\nu} 
  \biggl(\frac{h_{0}^2 \Omega_{\gamma0}}{2.47 \times 10^{-5}}\biggr),
\label{BBN1}
\end{equation}
where $\nu_{\mathrm{bbn}}$ and $\nu_{\mathrm{max}}$ are, respectively, the big-bang nucleosynthesis 
frequency and the maximal frequency of the spectrum. The bounds on $\Delta N_{\nu}$ range from $\Delta N_{\nu} \leq 0.2$ 
to $\Delta N_{\nu} \leq 1$. In the case $\Delta N_{\nu} < 1$ Eq. (\ref{BBN1}) would imply 
that the integrated spectral density is  about $10^{-5}$ and this bound is abundantly satisfied by Fig. \ref{Figure3}. 
\begin{figure}[!ht]
\begin{center}
      \epsfxsize = 11 cm  \epsffile{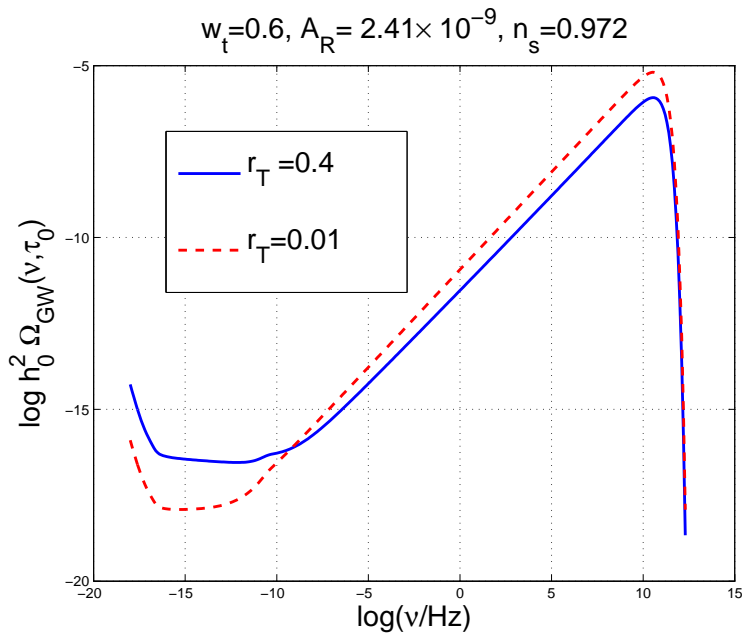}
\end{center}
\caption[a]{The cosmic background of relic gravitons is illustrated assuming that an absolute normalization is lacking. The barotropic index 
of the post-inflationary epoch is $w_{t} =0.6$.}
\label{Figure5}
\end{figure}
In the absence of an absolute normalization of the spectrum (such as the one provided by Bicep2) the constraints 
coming from low and high frequencies are qualitatively equivalent with the only difference that 
increasing spectra have to be preferentially constrained at high frequencies while 
decreasing spectra are more severely bounded at low frequencies. At the moment, however, 
we do not simply have a constraint at low frequencies but rather an explicit determination of $r_{T}$.
This observation imposes then an absolute normalization of any theoretical calculation of the cosmic graviton background, 
not only in the conventional case but also in the unconventional extensions of the standard lore. To illustrate this 
point it is useful to consider a specific case where the barotropic index of the post-inflationary expansion is given by $w_{t} \neq 1$ but 
the value of $r_{T}$ is left free to change in compliance with the consistency relations. For a generic $w_{t}$, the slope of 
$\Omega_{\mathrm{GW}}(\nu,\tau_{0})$ for $\nu> \nu_{\mathrm{s}}$ goes approximately as $\nu^{\delta}$  with $\delta = (6 w_{t} -2)/(3 w_{t} +1)$. For 
$w_{t} \to 1$ we also have $\delta\to 1$ and the case of the transfer function of Eq. (\ref{TT}) is recovered, up to logarithmic corrections. Of course, 
when $w_{t}\neq 1$ also the value of $\nu_{\mathrm{s}}$ will differ from the one of Eq. (\ref{FFs}). 
In Fig. \ref{Figure5} the normalization 
of the spectral energy density is imposed at high-frequency and this procedure is illustrated for the academic case $w_{t} = 0.6$. The dashed line in Fig. \ref{Figure5} leads to a promising signal at the scale of wide-band 
detectors but with $r_{T} =0.001$. Conversely the full line holds for $r_{T} =0.4$, i.e. twice the current Bicep2 determination. 
If we take the Bicep2 determination at face value we have that $r_{T}$ is not a tunable parameter and all the spectra 
must be normalized at low-frequencies in spite of their high-frequency behaviour.  
The direct limits coming from wide-band interferometers in their improved versions should therefore be considered in conjunction with the low-frequency determinations of $r_{T}$: in this way, they will provide extremely valuable informations on the cosmic background of relic gravitons. For instance the simple example of Fig. 
\ref{Figure5} shows that a sensitivity ${\mathcal O}(10^{-11})$ in $h_{0}^2 \Omega_{\mathrm{GW}}(\nu_{\mathrm{LV}}, \tau_{0})$ could potentially rule 
out directly the spectrum $w_{t} =0.6$ assuming an absolute normalization $r_{T} = {\mathcal O}(0.2)$. This would be, in turn,  a powerful constraint 
on the post-inflationary expansion rate.
\begin{figure}[!ht]
\begin{center}
      \epsfxsize = 11 cm  \epsffile{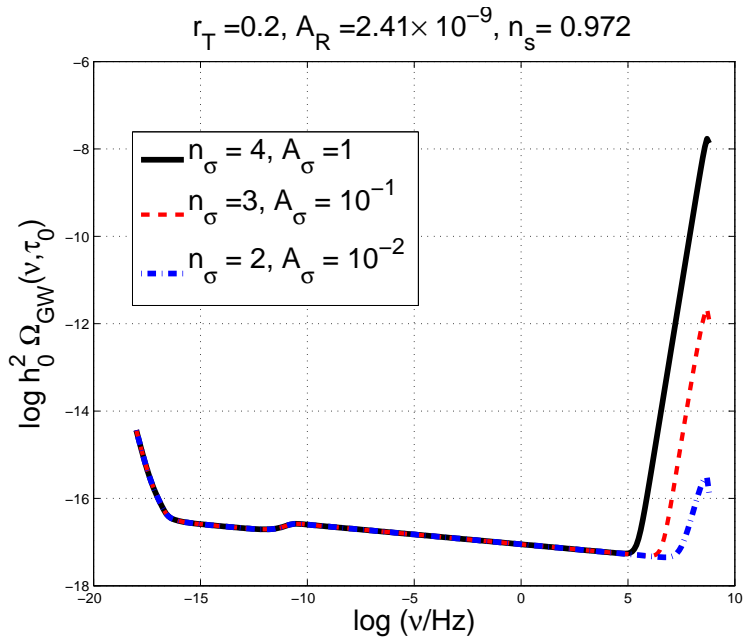}
\end{center}
\caption[a]{The cosmic background of relic gravitons  is illustrated for few specific choices of the 
amplitude and slope of the spectrum of the waterfall field.
The common logarithm is used on both axes. }
\label{Figure6}
\end{figure}

The results of Fig. \ref{Figure3} can then be used for comparing the theoretical signals with the sensitivities of wide-band 
interferometers such as LIGO/Virgo \cite{virgoligo}, TAMA \cite{TAMA} and Geo
\cite{GEO}. The sensitivity of a given pair of wide-band detectors to a stochastic background of relic gravitons depends upon 
the relative orientation of the instruments. The wideness of the band 
(important for the correlation among different instruments)
is not as large as $10$ kHz  but typically narrower. In an optimistic perspective,  it could range up to $100$ Hz.
 There are furthermore daring projects of wide-band detectors in space like the LISA \cite{lisa}, the BBO/DECIGO
\cite{BBODECIGO}. The common feature of these three projects 
is that they are all space-borne missions; the LISA interferometer should operate between $10^{-4}$ and $0.1$ Hz;
 the DECIGO project will be instead sensitive to frequencies between $0.1$ and $10$ Hz.

Using always the illustrative example of Fig. \ref{Figure3} we have that for $\nu={\mathcal O}(10^{-3})$ Hz (i.e. compatible 
with the frequency of space-borne missions) we would have $h_{0}^2 \Omega_{\mathrm{GW}}(\mathrm{mHz}, \tau_{0}) = 3.76 \times 
10^{-17}$ (including the effect of the running of the spectral index). For the Ligo/Virgo frequencies we would have instead 
\begin{equation}
8.56\times 10^{-17}\leq  h_{0}^2 \Omega_{\mathrm{GW}}(\nu_{\mathrm{LV}}, \tau_{0})\leq 7.10 \times 10^{-14}
\end{equation}
where the lowest value corresponds to $\nu_{\mathrm{LV}} =10$ Hz and the upper value 
corresponds to $\nu_{\mathrm{LV}} =10$ kHz. We shall not dwell here on the detectability prospects 
of other models that may offer higher signals in the Ligo/Virgo window \cite{max12}.
\subsection{Waterfall transitions}
The possibility of a long post-inflationary phase stiffer than radiation is not the only source of high-frequency 
indetermination of the cosmic graviton background. 
Another potential source of enhancement of the graviton background at high frequencies 
is provided by waterfall fields which a kind of spectator fields contributing 
to the transverse and traceless part of the total anisotropic stress \cite{mg2}. 
The two point function of the canonically normalized waterfall field is defined as
\begin{equation}
 \langle \sigma(\vec{k},\tau) \sigma(\vec{p},\tau) \rangle = \frac{2 \pi^2 }{k^3} {\mathcal P}_{\sigma}(k,\tau) \delta^{(3)}(\vec{k} + \vec{p}), 
\label{spect1}
\end{equation}
the power spectrum of $\sigma$ can be parametrized as\begin{equation}
{\mathcal P}_{\sigma}(k,\tau) = A_{\sigma}^2 \biggl(\frac{k}{k_{\mathrm{max}}}\biggr)^{n_{\sigma} -1} {\mathcal F}(k\tau),
\label{spect2}
\end{equation}
where $A_{\sigma}$ has the dimension of an inverse length and $n_{\sigma} $ is the spectral slope. In Eq. (\ref{spect2}) the dimensionless function ${\mathcal F}(k \tau)$ accounts for the time dependence  and the power spectrum is normalized at the comoving wavenumber $k_{\mathrm{max}}$ corresponding to the comoving frequency $\nu_{\max}$. If $n_{\sigma} > 3$ the spectral slope is steeper than  in the case of vacuum fluctuations. The amplified spectrum characterizing the waterfall field in hybrid inflation leads, according to recent analyses (see e.g. \cite{mg2} and references therein), to $n_{\sigma} \simeq 4$.

Since the waterfall field contributes to the anisotropic stress, the evolution of the tensor modes 
of the geometry of Eq. (\ref{mot1A}) must be complemented with the source term coming from the 
waterfall field which acts much earlier than the anisotropic stress produced by neutrino 
free streaming.  In Fig. \ref{Figure6} we illustrate the high-frequency enhancement of the spectrum for different 
spectral indices and different amplitudes $A_{\sigma}$ expressed in units of the reduced Planck mass $\overline{M}_{P}$ .

In the case of waterfall field the modification of the graviton background occurs for 
    frequencies larger than $10^{5}$ Hz  since this follows from simple estimates of the 
    width of the waterfall transition. In general we can say that the 
    turnaround frequency in the waterfall case is of the order of $e^{-N_{w}}$ times 
    $k_{\mathrm{max}}$. In this context $N_{w}$ is the number of efolds (between 4 and 5) 
    during the waterfall transition (see also \cite{wat2})
    
Comparing the spectra of Figs. \ref{Figure3}, \ref{Figure4} and  \ref{Figure6} we can appreciate different features that are not peculiar 
of the specific examples but have more general content. Typically the modifications of the post-inflationary history lead to 
spectra that are broader than in the waterfall case.
If the $\Lambda$CDM paradigm with $r_{T} = {\mathcal O}(0.2)$ is complemented by a high-frequency branch 
the maximal signal occurs in a frequency region between the MHz and the GHz. 
This intriguing aspect can be potentially scrutinized with
microwave cavities  \cite{HF1} or with other devices able to detect 
gravitational waves at high frequencies. Different groups are now concerned with high-frequency 
gravitons \cite{HF2,HF3}. It is not clear if, in the near future, the improvements in the terrestrial technologies will allow 
the detection of relic gravitons for frequencies, say, larger than the MHz.  At the same time 
relatively small interferometers, like the one under construction in Fermilab \cite{fermi}, may offer potential 
advantages in comparison with microwave cavities but in nearly the same frequency range. Unfortunately, however, 
the aim of the instrument does not seem to contemplate the direct search of the cosmic graviton backgrounds at high frequencies.

The illustrative examples of this section demonstrate that determination of $r_{T}$ does not 
exclude uncertainties at high frequencies. The strategy of normalizing the spectral energy density directly at high 
frequencies is now obsolete since the normalization is imposed at low frequency by the Bicep2 determination. 
\renewcommand{\theequation}{5.\arabic{equation}}
\setcounter{equation}{0}
\section{Concluding remarks}
\label{sec5}
Between the low-frequency radio waves and the  $\gamma$-rays there are, roughly, 22 decades in frequency. 
A similar frequency gap separates the relic gravitons probed by the CMB polarization experiments and the ones falling within the operating window of wide-band interferometers.  The aim of this paper has been to bridge this gap by analyzing the interplay between the cosmic graviton backgrounds of inflationary origin and the ongoing observations of the B-mode polarization of the CMB. In the concordance scenario
the value of   $r_{T}$ determines the cosmic graviton background not only at low and intermediate 
frequencies but also over much higher frequencies. In this situation the spectral energy density (in critical units) at the scale 
of the wide band interferometers is ${\mathcal O}(10^{-17})$ and it gets even smaller at higher 
frequencies. Taking the recent Bicep2 observations at face value, the minute signals obtained here are predictions of a 
specific model and not just upper (or lower) limits obtained from different 
theoretical assumptions, as we used to speculate in the past.

The consistency relations are not contradicted by the available data but they may be violated, for different reasons, at low and high frequencies. The models leading to a modified spectrum of the high-frequency gravitons must be compatible with the 
concordance paradigm in the low-frequency domain. The primary objective of the present analysis has been to present the accurate 
computation of the graviton spectrum in the concordance model over frequencies much higher than the pivot scale at which
CMB measurements are typically conducted.  We then considered the highest 
frequency domain and argued that there can be large quantitative uncertainties.
If the cosmic graviton background has a high-frequency component 
deviating from the concordance model, the low-frequency determinations of the tensor to scalar ratio and the 
high-frequency limits on the spectral energy density are not independent. We argued that, 
in the future, the high-frequency uncertainties could be eliminated by combining low-frequency polarization experiments and 
the direct determinations of the spectral energy density of the gravitons from wide-band interferometers.

\section*{Note Added}

After this paper has been submitted for publication, the Bicep2 results have been published (see first paper of Ref. \cite{bicep2}).
In the published version of the paper there are few modifications, as it usually happens. In particular the collaboration 
modified some of the plots in section IX of the paper and added a note after the conclusion.
Comparing the unpublished and the published versions of the paper the following 
conclusions can be drawn:
\begin{itemize}
\item{} since this paper was submitted new information on polarized dust emission has become available from the Planck experiment in four new papers 
\cite{ADD1,ADD2}; as noted by the Bicep2 collaboration these papers restrict their analyses
 to regions of the sky where Òsystematic uncertainties are small, and where the dust signal dominates total emission,Ó and that this excludes 
 $21$\% of the sky that includes the Bicep2 region.
\item{} the Bicep2 collaboration also quotes some more recent analyses \cite{ADD3} (see also \cite{emission} already quoted in the 
former version of this manuscript) where the polarized synchrotron emissions and the polarized dust emissions for typical 
frequencies ${\mathcal O}(150\, \mathrm{Hz})$;
\item{} the Bicep2 collaboration suggests that while the new developments ``do not offer 
definitive information on the level of dust contamination in our field, they do suggest that it may well be higher than any of the models considered" in the paper (verbatim from the first 
reference of \cite{bicep2});
\end{itemize}
To these statements we wish to add the following considerations:
\begin{itemize} 
\item{} as already swiftly mentioned when quoting Ref. \cite{emission} the spatial variation in the spectral 
index of the polarized synchrotron emission can account for at most $20$\% of the Bicep2 signal;
\item{} between the unpublished and the published version of the Bicep2 paper some of the 
foreground models have been dropped; in particular in the unpublished versions of the Bicep2 paper 
considers a data driven model (for short DDM) constructed ``using all publicly available information from Planck" (in the jargon DDM2 model).
\item{} this DDM2 model seems to correspond to some digitised version of a powerpoint presentation (see footnote 33 
of the Bicep2 preprint) that has been subsequently digitized and analyzed also by other authors (see e.g. second paper of \cite{ADD3}).
\end{itemize} 
To date  the only publicly available data on the foregrounds in the BICEP2 
region are the WMAP data. The best evidence for the BICEP2 signal not being a foreground and the best evidence 
for the foregrounds being a possible contaminant both cannot come, in such an important matter, from digitizing powerpoint 
presentations that were not intended to be used this way. 

Considering all these developments the situation it is fair to conclude (as it could be easily understood even without 
using the unreleased Planck data) that multifrequency measurements of the foregrounds are the key 
for a complete understanding of the foregrounds and of other competing signals (see e. g. \cite{ADD4}). 
It would seem wise to encourage a closer cooperation of the Planck and Bicep2 collaborations 
especially in the light of the fact that some researchers are simultaneously members of Bicep2 
and of Planck. 

Let us finally mention, as already stressed in the bulk of the present paper, that the value $r_{T} \simeq 0.2$ is largely 
illustrative. Different values of $r_{T}$ will be explicitly reflected in some numerical differences that will leave completely 
unaltered the spirit of the present analysis and its potential implications for wide-band interferometers and for 
other high-frequency devices\footnote{We note that after this paper has been submitted a preprint from the LIGO/Virgo collaboration 
appeared on the net \cite{ADD5}. The illustrative theoretical models of \cite{ADD5} do not include all the effects addressed in this paper but they also 
mention the possibility of a stiff post-inflationary phase that has been originally suggested in \cite{mg1} and further analyzed in \cite{max12}. 
If is unclear what are the large-scale data employed to implement the low-frequency normalization of the illustrated theoretical spectra. }.

\section*{Acknowledgments}

I thank G. Altarelli and for valuable exchanges of ideas.

\end{document}